\documentclass{article}

\usepackage{arxiv}

\usepackage[utf8]{inputenc} 
\usepackage[T1]{fontenc}    
\usepackage{hyperref}       
\usepackage{url}            
\usepackage{booktabs}       
\usepackage{amsfonts}       
\usepackage{nicefrac}       
\usepackage{microtype}      
\usepackage{lipsum}		
\usepackage{graphicx}
\usepackage{natbib}
\usepackage{doi}
\usepackage{multirow}%
\usepackage{amsmath,amssymb}%
\usepackage{amsthm}%
\usepackage{mathrsfs}%

\newcommand{\e}{\mathtt{e}}
\newcommand{\p}{\mathtt{p}}
\newcommand{\pset}{\mathbb{P}}
\newcommand{\id}{\mathtt{ib}}
\newcommand{\qb}{\mathtt{qb}}
\newcommand{\cep}{C_{\e,\p}}
\newcommand{\cip}{C^\id_{\text{clicks},\p}}
\newcommand{\cqp}{C^\qb_{\text{clicks},\p}}

\title{Analyzing News Engagement on Facebook: Tracking Ideological Segregation and News Quality in the Facebook URL Dataset}

\author{ \href{https://orcid.org/0000-0002-1647-7300}{\includegraphics[scale=0.06]{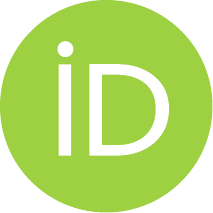}\hspace{1mm}Emma Fraxanet} \\
	Department of Engineering\\
	Universitat Pompeu Fabra\\
	Barcelona, Spain \\
	\texttt{emmafraxanet@gmail.com} \\
	\And
	\href{https://orcid.org/0000-0002-2271-3066}{\includegraphics[scale=0.06]{orcid.pdf}\hspace{1mm}Andreas Kaltenbrunner} \\
        Internet Interdisciplinary Institute (IN3) \\
	Universitat Oberta de Catalunya\\
	Barcelona, Spain \\
	\texttt{akaltenbrunner@uoc.edu} \\
    \And
	\href{https://orcid.org/0000-0002-6211-4519}{\includegraphics[scale=0.06]{orcid.pdf}\hspace{1mm}Fabrizio Germano} \\
	Department of Economics and Business\\
	Universitat Pompeu Fabra\\
	Barcelona, Spain \\
	\texttt{fabrizio.germano@upf.edu} \\
    \And
	\href{https://orcid.org/0000-0001-5146-7645}{\includegraphics[scale=0.06]{orcid.pdf}\hspace{1mm}Vicenç Gómez} \\
	Department of Engineering\\
	Universitat Pompeu Fabra\\
	Barcelona, Spain \\
	\texttt{vicen.gomez@upf.edu} \\
}

\renewcommand{\shorttitle}{Analyzing News Engagement on Facebook (Fraxanet et al.)}

\hypersetup{
pdftitle={A template for the arxiv style},
pdfsubject={q-bio.NC, q-bio.QM},
pdfauthor={David S.~Hippocampus, Elias D.~Striatum},
pdfkeywords={First keyword, Second keyword, More},
}

\begin{document}
\maketitle

\begin{abstract}
The Facebook Privacy-Protected Full URLs Dataset was released to enable independent, academic research on the impact of Facebook’s platform on society while ensuring user privacy. The dataset has been used in several studies to analyze the relationship between social media engagement and societal issues such as misinformation, polarization, and the quality of consumed news. In this paper, we conduct a comprehensive analysis of the engagement with popular news domains, covering four years from January 2017 to December 2020, with a focus on user engagement metrics related to news URLs in the U.S. By incorporating the ideological alignment and composite score of quality and reliability of news sources, along with users' political preferences, we construct weighted averages of ideology and quality of news consumption for liberal, conservative, and moderate audiences.
This allows us to track the evolution of (i) the ideological gap in news consumption between liberals and conservatives and (ii) the average quality of each group's news consumption. 
We identify two major shifts in trends, each tied to engagement changes. In both, the ideological gap widens and news quality declines. However, engagement rises in the first shift but falls in the second.
Finally, we contextualize these trends by linking them to two major Facebook News Feed updates. 
Our findings provide empirical evidence to better understand user behavior and engagement with news and their leaning and reliability during the period covered by the dataset.
\end{abstract}


\section{Introduction}\label{sec1}

The interplay between user ideology, news source quality, and online engagement has been a key focus of research in recent years. Studies consistently show that individuals tend to consume ideologically aligned content, with less moderate views often linked to lower-quality news sources~\cite{bakshy2015exposure, robertson2023users}, and more extreme narratives being associated with higher engagement~\cite{galeazzi2024unveiling}. Extreme conservative audiences, in particular, have been found to exhibit higher levels of ideological segregation~\cite{gonzalez2023asymmetric, heseltine2023asymmetric}. These patterns raise concerns about the relationship between online platforms' design and the segregation of information consumption, as platform-specific mechanisms can influence the extent of ideological separation.
For instance, ideological segregation appears stronger on Facebook than on Reddit, where interactions remain more heterogeneous even among extreme-leaning users~\cite{cinelli2021echo}. Moreover, selective exposure to partisan news sources tends to increase with user activity~\cite{cinelli2020selective}, contributing to the formation of highly clustered news consumption patterns~\cite{schmidt2017anatomy}. 


Recent studies examining social media engagement, particularly around the 2020 US Election, have highlighted the significant impact of algorithmic features on user experience and the non-trivial correlation they may have on polarization~\cite{garcia2023influence}. For instance, experiments altering algorithmic feed rankings revealed that chronological feeds may increase exposure to untrustworthy political content while decreasing user engagement and satisfaction~\cite{guess2023social}. Similarly, efforts to reduce exposure to like-minded sources showed limited efficacy in addressing polarization~\cite{nyhan2023like}. Moreover, studies have found evidence that re-shares —the feature facilitating \textit{virality}— may not be the only driving factor generating misinformation and polarization~\cite{guess2023reshares}.

Building on the literature on algorithmic amplification, recommender systems have been shown to prioritize engagement, often at the expense of exposing users to more polarizing or toxic content, thereby amplifying ideological divides~\cite{chavalarias2024can}. Regarding the assessment of partisan audience bias in social media,  \citet{robertson2018auditing,robertson2023users} have conducted audits and investigations on user exposure and interactions with content on Google Search, while \citet{bakshy2015exposure} used US Facebook data on interactions with shared news and friend networks. Contrarily to \citet{gonzalez2023asymmetric}, both concluded that individual choices played a stronger role in increasing like-minded content, as well as limiting cross-cutting exposure than algorithmic ranking. Additionally, Gentzkow and Shapiro \cite{gentzkow2011ideological} and Flaxman et al. \cite{flaxman2016filter} showed that while ideological filtering existed online, it was partly offset by incidental exposure to diverse viewpoints, particularly through social media These contrasting findings suggest the need for a more comprehensive analysis of user behaviors and algorithmic influences~\cite{budak2024misunderstanding}.

Despite these advances, much of the existing research relies on short time frames or experiments done on small treatment groups that are not completely isolated from other platform-wide dynamics~\cite{gonzalez2023asymmetric}. Moreover, most findings reflect specific moments in a platform's evolution, often ignoring the ongoing changes to algorithmic features, UX design, privacy settings, or moderation strategies. These changes are particularly problematic because they are rarely disclosed in full. In the few cases where some information is available, the timing and specifics of algorithmic modifications remain opaque, even to researchers conducting preregistered experiments in collaboration with the platform. For instance, the ``break-glass'' changes to the news feed on Facebook in 2020, which coincided with the study by \citet{guess2023social}, illustrate how such undisclosed adjustments can potentially affect the conclusions of otherwise well-designed research~\cite{bagchi2024social} (see also the e-letters below \cite{guess2023social} for more details on this discussion).

Therefore, while these studies are very informative on the behavior of different population segments in the online realm, they usually cannot address the interplay of these behaviors and platform characteristics. Understanding whether the behaviors are stable through time or, on the other hand, subject to both exogenous and endogenous changes is relevant for promoting healthier online spaces.

In this paper, we address this problem through an aggregate, longitudinal analysis of engagement patterns, leveraging the temporal characteristics of the Facebook URL dataset. Our analysis identifies key change points that align with major updates reported in the News Feed Facebook's algorithm.

The \textit{Facebook Privacy Protected Full URLs Dataset}~\cite{messing2020facebook} is a large-scale collection of public posts containing URLs and user interaction data on the Facebook platform starting January 2017. The dataset includes counts of engagement at the URL-action level, stratified by limited user attributes such as political page affinity (PPA), age group, and gender. It does not contain, however, information on the volume of unique users in each of these categories or in the dataset in general. While the dataset is unique in its size and granularity, it also includes privacy-protection mechanisms such as injected noise and a minimum-share inclusion threshold that introduce certain limitations. We discuss these in more detail in Section~\ref{sec:dataset_issues}.

Despite these limitations, researchers have used the Facebook Privacy-Protected Full URLs Dataset for various applications. Our work, akin to \citet{guess2021cracking} and \citet{bailey2021interactions}, provides a descriptive static analysis of exposure and sharing patterns of news URLs in the US, confirming findings like the conservative skew in low-quality news consumption.
Building on this foundation, our work introduces a temporal dimension to investigate ideological segregation and information quality in news consumption. Similar to \citet{buntain2023measuring}, who developed a metric for robust domain ideology scores based on exposure and engagement, we focus on user behavior and content interactions but expand the scope by examining longitudinal trends and dynamic patterns over time. 

Prior work on misinformation has shown that platform interventions, such as labeling or demoting low-quality pages, have limited long-term effects~\cite{thero2022investigating, vincent2022measuring}.
\citet{bandy2023facebook} show that Facebook amplified low-quality publishers during the 2020 US election, with algorithmic changes impacting publishers broadly, regardless of content quality. We extend this line of inquiry by contextualizing engagement trends within potential algorithmic adjustments \cite{meta_2018, wsj_2021, narayanan2023smalg} to examine their implications for ideological segregation on the platform.

Our study contributes specifically to the literature on \textit{news audience polarization}~\cite{mangold2022metrics}, focusing on ideological segregation in engagement on Facebook (i.e. how liberal and conservative users differ in their interactions with news sources over time). Using a four-year longitudinal dataset of Facebook interactions, we characterize engagement trends (e.g., clicks, likes, shares, and comments) across news-related domains. We characterize these patterns in relation to user ideology, domain ideology, and domain quality.  By developing a novel metric to assess partisan bias in user consumption patterns, we track the evolution of the ideological gap between news consumed by conservatives and liberals as an indicator for polarization in news diets.

While broader debates on political polarization encompass ideological, affective, and partisan dimensions~\cite{baldassarri2008partisans}, and findings on the role of online platforms remain mixed~\cite{flaxman2016filter, gentzkow2011ideological, guess2021almost, boxell2017internet, allcott2024effects}, our analysis does not aim to assess polarization in society at large. Instead, we examine group-level engagement patterns specific to Facebook, shaped by both user preferences and platform design using a longitudinal dataset.

Additionally, we study changes in the quality of news consumption and how it relates to both user and outlet ideology. To do so, we examine content trustworthiness and integrity using a domain-level quality score based on expert evaluations from Lin et al.~\cite{lin2023high}, which captures both journalistic standards and a domain's history of sharing misleading content.  While a low-quality score does not equate to the presence of misinformation in every piece of content from that outlet, we interpret low domain quality as indicative of a higher likelihood of such behavior, while avoiding content-level claims. Therefore, our use of the term “misinformation” throughout this work refers to a broad behavioral pattern of spreading false or misleading content, and not the narrower, intent-based definition that distinguishes disinformation.

In particular, we address the following research questions in our aggregate statistical analysis:
\begin{description}
\item[\textbf{RQ1}:] \emph{How do engagement patterns on Facebook differ between popular news and other URLs, and how do they vary by user ideology?}
\item[\textbf{RQ2}:] \emph{For news-related URLs, how are users’ political page affinities related to the orientation and quality of the domains they engage with?}
\end{description}
while in our longitudinal analysis over a four-year period, we address the following:
\begin{description}
    \item[\textbf{RQ3}:] \emph{How do engagement patterns with news-related URLs evolve over time, and what are the key change points that mark significant shifts in user engagement?}
    \item[\textbf{RQ4}:] \emph{How such engagement patterns differ from other URLs?}
\end{description}
while for news-related URLs we investigate:
\begin{description}
    \item[\textbf{RQ5a}:] \emph{How does ideological segregation in news consumption change over time?}
    \item[\textbf{RQ5b}:] \emph{How does the consumption of low-quality news vary over time?}
\end{description}

Our findings provide a detailed view of the dynamic relationship between user behavior, news quality, and ideological segregation in news diets, while situating these changes within the broader context of platform-level adjustments and potential algorithmic modifications.

\section{Data Collection, Processing, and Descriptive Analysis}

The Facebook URL Dataset allows queries at the URL-action level including the numbers of views, clicks, likes, shares, comments, and emoji reactions (\textit{angers}, \textit{hahas}, \textit{wows}, \textit{loves} and \textit{sorrys}). Views and clicks are typically regarded as passive forms of engagement or consumption (used interchangeably here), while the other metrics represent active engagement. The counts are also broken down by month and audience demographics (country, age, and gender). 

For the US, these counts are also stratified by the estimated ideological leaning of the users, called \textit{Political Page Affinity} or PPA, which classifies the audience into five different buckets, scaled from $-2$ (most liberal) to $+2$ (most conservative), with an additional bin (not defined). This metric, based on the model described in~\citet{barbera2015birds}, is computed internally at Facebook. For that reason, we focus on URLs shared primarily in the US and extract engagement counts from users located in the US only.

\subsection{Domain and Data Selection Criteria}
\label{sec:domains}

For this study, we separate between: (i) news-related domains with reliable annotations for both ideological leaning and quality, and (ii) other domains, that are potentially non-news and include all remaining sites in the Facebook URL dataset, such as entertainment, e-commerce, or low-engagement domains, for which ideological or reliability classifications are not systematically available.

Our selection of news-related domains is conceptually grounded in our goal to examine ideological segregation and information quality in news consumption. As such, we include only domains that (1) are among the top 1\% most-viewed domains in the U.S. Facebook dataset (2017–2021), and (2) have both domain-level ideology and quality scores. This filtering is necessary to ensure that each domain in our analysis can be meaningfully positioned along both dimensions.

We construct the set of news domains using the following three sources:

\begin{enumerate}
    \item \textbf{Top 1\% most-viewed domains}. We begin with the list of the $2,629$ most-viewed domains~\cite{buntain2023measuring}, which account for over 77\% of all URL views and over 80\% of all URL clicks in the U.S. on Facebook from 2017 to 2020. Given the heavy-tailed distribution of engagement on social media, this threshold is not arbitrary but reflects a pragmatic balance between coverage and signal quality. The domains in this selection range from very high view counts per month (on the order of $10^{11}$) to lower values (of the order of $10^{3}$), reflecting the wide engagement spectrum within the top 1\%. Including domains with even lower engagement would introduce substantial noise due to differential privacy constraints, as such domains tend to hover near the dataset’s inclusion threshold and offer a limited usable signal.
    \item \textbf{Domain quality scores}. 
     We use the domain-level \textit{quality scores} from~\citet{lin2023high}, who define a composite quality score based on evaluations from six independent expert rating sources (including NewsGuard\footnote{\url{https://www.newsguardtech.com/}}). These scores reflect a broad range of criteria—such as factual reporting, transparency, editorial standards, and importantly, the frequency with which an outlet has been known to share false or misleading content. The composite score, derived via principal component analysis, captures a latent “domain reputation” dimension that blends aspects of both journalistic professionalism and misinformation-related behaviors. 
     The substantive meaning of the PCA score is not directly interpretable in terms of any one dimension of quality, since it reflects the dominant axis of variation in the rating dataset, which may prioritize some attributes (e.g., factual accuracy) over others (e.g., transparency of sourcing) depending on their variance.   
     We acknowledge that in media and communication studies, the term “quality” often has a restricted normative meaning tied to journalistic professionalism. 
     Our use of the term adopts a broader empirical framing that includes reputational and factual indicators, including a domain's historical association with false or misleading content. 
     Therefore, a low score may reflect various factors, such as misinformation prevalence or low transparency, without distinguishing between them. Quality scores are available for 1,586 ($\approx 63\%$) of the $2,629$ domains.
    \item \textbf{Ideology scores}. We use the domain-level ideology scores from~\citet{robertson2018auditing}, who developed a domain-level audience bias metric based on the sharing behavior of registered US voters on Twitter. These scores reflect aggregate partisan preferences in domain sharing and offer a scalable, behavior-based measure of ideological orientation. Compared to manual annotation or content-based scoring, this method avoids rater subjectivity and better captures the political leanings of domains as experienced by users. The metric has been validated against several external sources, showing high agreement with prior audience-based and rater-based measures, and is widely used in polarization and media bias research. These scores are continuous, behavior-based, and include $1,664$ of the $2,629$ domains ($\approx 63\%$). 
\end{enumerate}

After intersecting these three sources, we identify 1,231 domains as news-related. This set includes a wide range of outlets: mainstream, local, partisan, centrist, and fringe, and represents a broad and ideologically diverse sample. Among them, approximately 58\% have a quality score greater than 0.6, a threshold aligned with NewsGuard’s definition of high trustworthiness~\cite{lin2023high}. It is important to clarify that the remaining 53\% of domains from the original top 1\% list are not necessarily excluded news domains. Many are non-news sites (e.g., memes, e-commerce, clickbait farms), and are excluded because they lack ideological or quality annotations. See Table~\ref{t:stats} for the number of domains and related URLs for each of the two categories.

\begin{figure}[h!]
\centering
\includegraphics[width=0.5\textwidth]{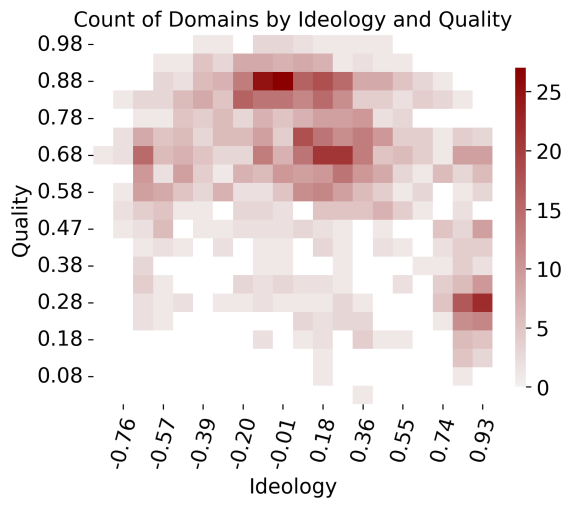} 
\caption{\textbf{Selected news domains for this study distributed by Ideology and Quality scores.} (bin width ideology $= 0.094$, quality $= 0.05$, ticks mark the center of the bin) Positive values of ideology denote a leaning toward conservative ideology, while negative values correspond to liberal leanings. Quality scores are higher when domains have higher quality. The two dimensions correlate negatively ($r = -0.35, p < 0.001$). }
\label{fig:domains}
\end{figure}

As Figure~\ref{fig:domains} shows, there is a weak negative correlation between domain ideology and domain quality scores ($r = -0.35, p < 0.001$), with two clusters of high and medium-high reliability domains located near the ideological center and a concentration of low-quality domains toward the conservative end of the spectrum. Additionally, when considering ideological extremity (i.e., the absolute value of ideology scores), we observe a somewhat stronger negative correlation ($r = -0.48, p < 0.001$), indicating that lower quality could be associated with more ideologically extreme domains, regardless of direction.

\begin{table}[]
\caption{\textbf{Summary of Domain and URL statistics for popular news domains and other domains.}}
\begin{tabular}{l|r|r}
& \multicolumn{1}{c|}{News} & Other \\
\hline
Total domains         &     1,231 &   268,821 \\
Total number of URLs  & 6,273,432 & 5,215,067 \\
URLs per month        &   130,697 &   108,647 \\
Stdev. URLs per month &    42,145 &    38,541
\end{tabular}
\label{t:stats}
\end{table}

\subsection{URL Retrieval and Preprocessing}\label{sec:URLs}

We retrieve URLs through successive monthly queries to the platform, covering the period from January 2017 to December 2020. The dataset is protected by differential privacy~\cite{messing2020facebook, evans2023statistically}, which prevents tracing actions to specific users or URLs. However, while differential privacy safeguards data, it can introduce biases into inference methods if not properly accounted for~\cite{evans2023statistically}. Additional challenges of this dataset include selection bias caused by engagement thresholds for URL inclusion~\cite{allen2021research} and past data integrity issues~\cite{allen2022addendum}. An example of a precise query with parameters is detailed in \textit{Supplementary Fig. S1}, and a discussion of these limitations and our strategies for addressing them can be found in Section~\ref{sec:dataset_issues}.

The inclusion of URLs in the dataset is cumulative: new URLs are incorporated every month if they meet the minimum of 100 shares (plus Laplacian noise) threshold, but they are never removed~\footnote{Some retrieved URLs reported engagement data up to one month prior to their posting date (approximately $9000$ cases for 2019) or even earlier (around $200$ cases across all years). Engagement from the former group was aggregated into the zero-month time difference, as it typically corresponds to activity within hours before the month begins. The latter cases were excluded from the analysis.} once this threshold has been surpassed at least once. 

\begin{figure}[h!]
\centering
\includegraphics[width=0.95\textwidth]{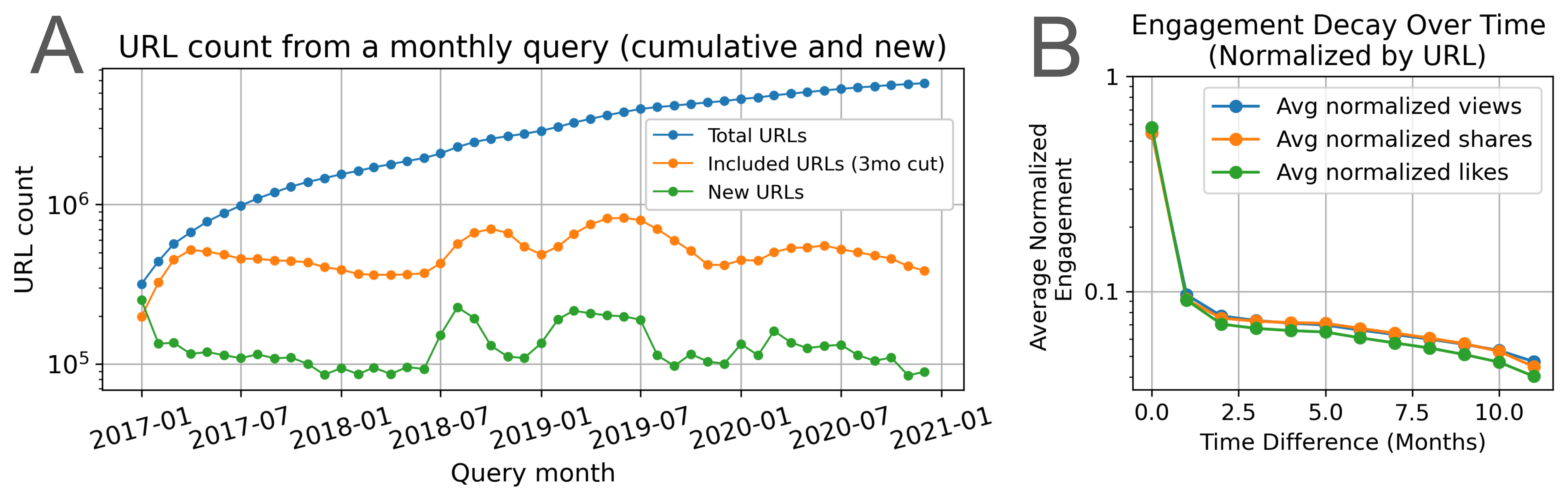}
\caption{\textbf{(A) Volume of URLs per query and analysis of engagement decay per URL (for news domains).} In blue, the total number of URLs retrieved per each monthly query. In orange, the ones considering a 3-month limit after each URL's posting date (the ones used for the analysis). In green, the number of new URLs added to the dataset each month. \textbf{(B) Average (normalized) engagement}. Engagement decays fast, and most of the engagement occurs during the first few months, on average. While this figure specifically represents data from 2019, similar patterns are observed across all analyzed years. }
\label{fig:nurls}
\end{figure}

Figure~\ref{fig:nurls}A shows the number of unique URLs retrieved per month for news domains. The total number of URLs (in blue) reflects the cumulative count over time. A subset of them (count in green) only includes the new URLs added each month, which amounts to between \(9 \cdot 10^4\) and \(2 \cdot 10^5\) URLs. 
For efficiency, we discard URLs that appeared in the dataset before a cutoff period preceding the queried month. We set the value of this cutoff to \emph{three months} for reasons explained below. Therefore, the analysis for each month includes only URLs that were incorporated during the queried month or within the preceding three months (shown in orange).

To justify our choice of a cutoff period of three months, we calculate the typical lifespan of a URL. Figure~\ref{fig:nurls}B shows how engagement decays (on average) over time for the year 2019 (all years behave similarly) for three different types of engagement. We compute the total engagement for a URL within that year and normalize the proportion of engagement for each month, starting from the month in which it was first posted and continuing with subsequent months. The curve illustrates the expected lifespan of a URL's engagement once it appears in the dataset. We observe that most of the engagement ($\approx 70\%$) occurs during the first two months. We repeated \emph{all our analyses} using different values for the cutoff period (3-month, 6-month, and 12-month cutoffs) and found that the results for Figures 3 through 7 remained identical across these settings. Because the figures are visually indistinguishable, this data is not shown. This indicates that extending the cutoff period essentially increases the size of the monthly URL set (see~\ref{fig:nurls}A), but the overall impact on engagement is negligible. Consequently, we present our results considering engagement counts using a 3-month cutoff.

\section{Aggregate Analysis of Engagement}

\subsection{RQ1. Engagement Patterns by User Ideology}

First, we analyze the distribution of engagement across user PPA categories. Since the dataset is designed to include noisy aggregated interaction counts per PPA (without information about the number of unique users per interaction), we experimented with different baselines and normalization methods. Details on the methods used to compute confidence intervals for engagement counts can be found in the \textit{Supplementary Information 1A}.

Given the aggregated counts $\cep$ for each engagement type $\e\in\{\text{clicks}, \text{views}, \hdots\}$ and user political page affinity $\p\in\{-2,-1,0,1,2\}$, we compute the (normalized) engagement~$E_{\e,\p}$ relative to a baseline of views as
\begin{align}\label{eq:counts}
    E_{\e,\p} &= \frac{\bar{C}_{\e,\p}}{\displaystyle\max_{\p'}  \bar{C}_{\e,\p'}}, &
    \bar{C}_{\e,\p} & =\frac{C_{\e,\p}}{C_{\text{views},\p}}.
\end{align}

Normalizing by views captures the engagement given the exposure to the content and normalizing by the maximum value allows to compare across different types of engagement, despite their varying volumes. Alternative normalization methods and resulting outcomes are discussed in the \textit{Supplementary Information 2}.

\begin{figure*}[t!]
\centering
\includegraphics[width=0.95\textwidth]{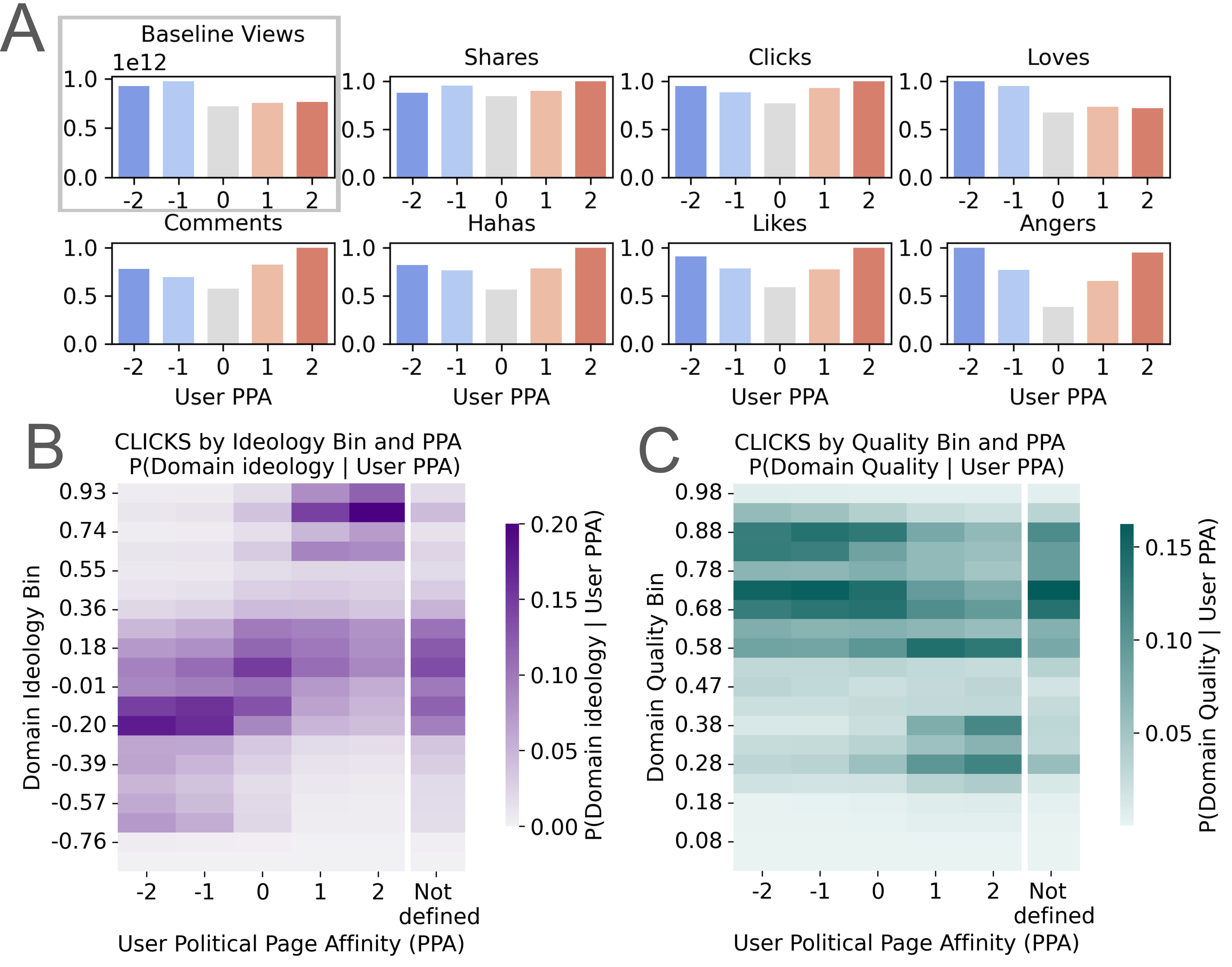}
\caption{\textbf{Aggregate Analysis of Engagement (for popular news domains).}
In all figures, negative values correspond to more liberal-leaning, and positive values correspond to more conservative-leaning. 
\textbf{(A) Distribution of different engagement types across user political page affinity (PPA)}, normalized relative to view counts (top left). For engagement types other than views, values are further normalized relative to their maximum across the five PPA categories, see Eq.~\eqref{eq:counts}. Engagement is unevenly distributed, with liberals having significantly more views than conservatives. The U-shaped pattern of engagement (greater engagement by extreme users) becomes more pronounced from left to right and from top to bottom. Users with undefined PPA exhibit much higher passive engagement (views and clicks) but significantly fewer engagement counts per view and are therefore not represented.
\textbf{(B) Conditional probability of clicking on a URL with a specific ideology given the user PPA.} (bin width $=0.094$, columns sum one). Liberal and center users follow single-mode distributions of engagement, with liberals leaning towards negative (liberal) domain scores and center users favoring neutral domains. Conservative users display a bimodal distribution, engaging both with neutral and far-right domains. Undefined users exhibit a profile similar to center users.
\textbf{(C) Conditional probability of clicking on a URL with a specific quality given the user PPA.} (bin width $=0.05$, columns sum one). Liberal users predominantly engage with high-quality domains. Center and conservative users progressively interact with lower-quality domains, with extreme conservatives showing the highest engagement with low-quality domains. This trend is supported by a negative correlation between domain quality and user ideology. See main text for details.}
\label{fig:aggregate}
\end{figure*}

Figure~\ref{fig:aggregate}A shows the results for popular news domains.  Engagement is unevenly distributed, with liberals having significantly more views than conservatives. 
This suggests that liberals tend to engage more frequently in passive consumption of news-related URL-containing posts, which we find to be opposite to other content (see \textit{Supplementary Fig. S3}).  
For clicks, the other type of passive engagement, the pattern shifts slightly to a U-shape, with center users having the lowest number of clicks. As for the active engagement counts, there is a clear U-shaped pattern for \textit{hahas}, likes, comments, and \textit{angers} (increasingly pronounced in that same order). Surprisingly, the distribution for shares is the least U-shaped of all, with other reactions such as \textit{loves} also having a flatter shape. We find a very similar behavior for engagement with other domains (\textit{Supplementary Fig. S3}).

Users with non-defined PPA (data not shown) have a view count that is three times higher but have similar values of active engagement in total. While the number of users in each PPA category is not included in the dataset, if we assume that the order of magnitude of views is informative of the number of users included in that category, this would imply that there is a large set of politically disconnected users that consume content passively but are much less active than politically categorized users when it comes to engaging with news-related posts. This lower activity may in part be the reason why they are not assigned to an ideology bin.

\subsection{RQ2. Domain Ideology and Domain Quality by User Ideology}

We now examine patterns in news consumption across different user PPA categories based on both the ideology and quality of the URLs. We use the ideology and quality scores introduced in Section~\ref{sec:domains}, which can be obtained for each individual domain. 
For clarity, we discretize the scores into bins.

We present results based on clicks since clicks are a form of passive engagement that is a more reliable indicator of consumption of the information contained in the URLs. A view could merely account for a loaded and quickly scrolled-over post in a user's feed. However, we studied the same patterns with several engagement metrics (views, clicks, shares, likes, and comments) and found no relevant differences between them.

Let $C^{\id}_{\text{clicks},\p}$ represent the sum of engagement counts on clicks from a user PPA category $\p$, for URLs whose domain ideology falls within domain ideology bin $\id$. We compute the conditional probability distribution of domain ideology given the user PPA category as $$P(i(u)\in\id|\p) =\cip/\sum_{\id'} C^{\id'}_{\text{clicks},\p} \quad.$$ In other words: given the PPA $\p$ of a user, and that this user has clicked on URL $u$, what is the probability that the ideology of URL $u$, denoted as $i(u)$, is in the ideology interval $\id$?  
Similarly, we compute the probability of clicking on a URL of domain quality $$P(q(u)\in\qb|\p) =\cqp/\sum_{\qb'} C^{\qb'}_{\text{clicks},\p} \quad,$$ where $\qb$ denotes a domain quality bin and $q(u)$ the domain quality of URL $u$ (see Section~\ref{sec:iwa} for a formal definition of these variables). 

Figure~\ref{fig:aggregate}B shows $P(i(u)\in \id|\p)$ as a heatmap. We see that, while users in the liberal and center classes seem to follow a single-mode distribution, with its center tilted towards negative values for liberals and centered for the center category, conservative users follow a bimodal distribution. We can see one mode at the center and another mode on the far conservative part (upper part in the heatmap). 

In Figure~\ref{fig:aggregate}C we show $P(q(u)\in\qb|\p)$. We observe that liberal-leaning users stay within the high-quality bins, while there is a slow progression towards wider distributions for the center and conservative-leaning users that eventually reach low-quality domains for the extreme conservatives. 
This result is in agreement with the observed relationship between domain ideology and domain quality at the URL-domain level, independent of user PPA (see Figure~\ref{fig:domains}).

Users with undefined ideology (referred to as ``Not defined'' in the figures) show engagement patterns similar to center users. This is supported by the Total Variation distance between the distribution of ``Not defined'' users and each ideological group: \{-2: 0.33, -1: 0.26, \textbf{0: 0.06}, 1: 0.31, 2: 0.41\} for ideology-based engagement (Figure~\ref{fig:aggregate}B), and \{-2: 0.10, -1: 0.09, \textbf{0: 0.08}, 1: 0.22, 2: 0.31\} for quality-based engagement (Figure~\ref{fig:aggregate}C). In both cases, ``Not defined'' users are most similar to center users in their engagement patterns.

\section{Longitudinal Analysis}

\subsection{RQ3 and RQ4. Engagement Timeline and Change Point Detection}

\begin{figure}[ht!]
\centering
\includegraphics[width=1\textwidth]{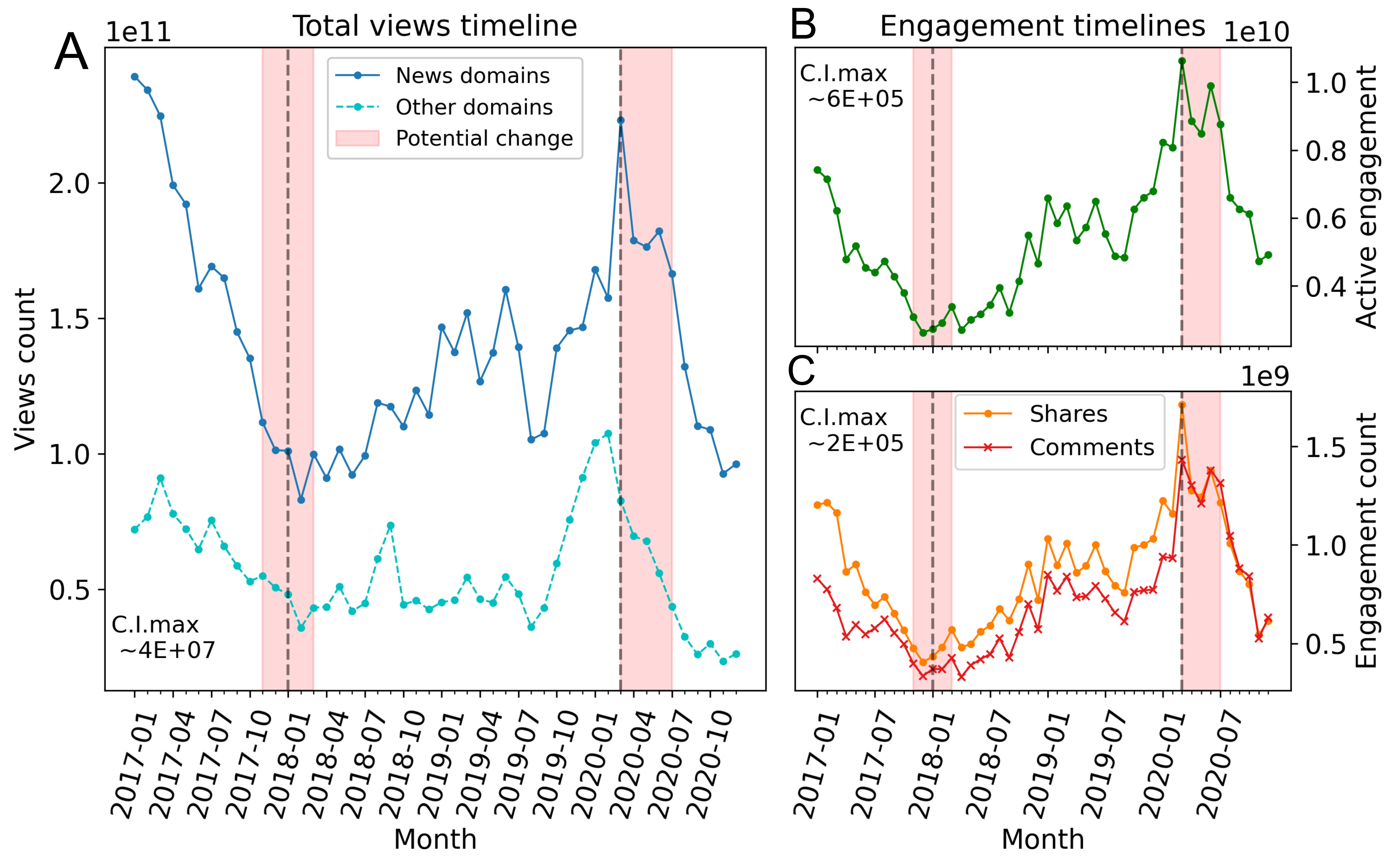} 
\caption{\textbf{Engagement timeline.}
\textbf{(A)} Total monthly counts of views for news and other domains (passive engagement); \textbf{(B)} aggregate of active engagement metrics for news domains (e.g., likes, shares, comments). \textbf{(C)} Two specific counts: shares and comments, also for news domains. Text labels indicate the maximum values of the noise uncertainty confidence intervals (CI), calculated using Eq.1 in \textit{Supplementary Information 2A}. Shaded regions represent potential change points, identified through piecewise linear regression. Dashed lines mark the dates of two major documented algorithmic changes.}
\label{fig:engagement}
\end{figure}

We begin this analysis by showing in Figure~\ref{fig:engagement} the time series data of raw engagement per month. The left plot displays view counts, the right upper plot shows aggregated active engagement, and the right bottom plot breaks down engagement by shares and comments. A figure illustrating all active engagement timelines is provided in \textit{Supplementary Fig. S4}.

The global temporal pattern for news domains is clear: a steady decline in engagement until the end of 2017, followed by a rise that continues until the first quarter of 2020, after which engagement abruptly declines again.

To quantify these shifts, we identify significant change points by applying segmented linear regression~\cite{muggeo2003estimating,pwlf} to each individual engagement timeline. Segmented linear regression models the data as a piecewise linear trend, aligning with the observed data and allowing for changes at distinct change points identified as change points. For most individual engagement metrics, the identified number of change points is two. 

Because our temporal resolution is monthly, and because different engagement metrics may respond to change at slightly different rates, we expect some small variation in the precise change point locations across engagement metrics. Instead of treating individual change points in isolation, we apply segmented linear regression independently on multiple time series (representing different engagement metrics) and then aggregate these results to identify broader change regions (i.e., bands of months where multiple metrics indicate potential structural shifts). These regions are marked by shaded regions in Figure~\ref{fig:engagement} and throughout other figures in the manuscript (see the \textit{Supplementary Information 4} for more details on the methodology). 

We complement these regions with dashed lines indicating the month when actual changes to the Facebook ranking algorithm were reported or expected to have occurred (see \cite{meta_2018, wsj_2021,narayanan2023smalg} for the 2018 update), which is associated with precise date; see \cite{narayanan2023smalg, dell2023fbpapers} for the 2020 update, for which we could not find a precise date but chose March as the most likely month in which it occurred (February and April could not be excluded). While the change points coincide with algorithmic changes, the extent to which these events are driven by the ranking algorithm or other exogenous factors cannot be determined from this analysis. We discuss this in more detail in Section~\ref{sec:context}.

In addition to these major trends, we highlight other interesting observations. For example, when comparing shares and comments, the absolute sum of shares (bottom plot, orange line) declines until it matches the absolute sum of comments (red line). This observation may be associated with a decrease in the weight of shares affecting feed ranking in 2020, as explained in the Discussion section.

We also show the time series of view counts for the domain URLs that are not in our news domains list in the left plot (A) of Figure~\ref{fig:engagement} (bottom curve). The general trend is very similar to news domains, which suggests that the detected change points apply platform-wide and are not exclusive to news-related content.
We can also observe three differences: (I) The total volume of views is always lower than for popular news domains, (II) The initial decrease seems to be less strong, and (III) the increasing trend between the two identified change point regions is less steady and more sudden. When separating active engagement for each different action, we find similar differences (\textit{Supplementary Fig. S4}). 

We conclude this initial analysis by noting that these trends correspond to the raw engagement time series. For example, engagement time series could be normalized by views to reduce common effects across all engagement types. The results of this normalization are shown in the \textit{Supplementary Information 5A, Fig. S5}, where we observe that the share of active engagement steadily increases over time compared to passive engagement.

\subsection{Measuring Segregation and Low-Quality News Consumption Over Time}

\label{sec:iwa} 
Our proposed methodology for calculating partisan segregation and the quality of news consumption from URL data relies on weighted averages, with the weights determined by the engagement counts associated with each URL.

For clarity, we merge moderate and strong user political affinities (PPA) into a combined political category $\pset$  resulting in four distinct combined categories: conservatives $\mathcal{C}$ (buckets $+2$ and $+1$), liberals $\mathcal{L}$ (buckets $-2$ and $-1$), centrists $\mathcal{N}$ (including only bucket $0$), and undefined~$\mathcal{D}$ (when the PPA is not defined). Merging strong and moderate buckets does not affect the results obtained.

Formally, let $u$ denote a URL present in the data (we omit the month index to simplify notation) and let $i(u) \in [0,1]$ and $q(u)\in[0,1]$ denote the domain-level ideology and domain-level quality scores of $u$, respectively~\footnote{To prevent numerical problems when calculating confidence intervals for the denominators, we normalize the domain ideology scores from $0$ to $1$, where $0$ indicates the most liberal-leaning domains and $1$ represents the most conservative-leaning domains in our dataset.}.

To compute the overall ideology and quality for engagement type $\e$ and combined user category $\pset$, we calculate the following weighted averages:
\begin{align}
    \begin{array}{c}
    \text{Domain}\\\text{Ideology}
\end{array}~~
    \mu^I_{\e,\pset} &= \frac{\sum_{u=1}^n i(u) C_{\e,\pset}^{(u)}}{\sum_{u=1}^n C_{\e,\pset}^{(u)}}, &
    \begin{array}{c}
    \text{Domain}\\\text{Quality}
    \end{array}~~
    \mu^Q_{\e,\pset} &= \frac{\sum_{u=1}^n q(u) C_{\e,\pset}^{(u)}}{\sum_{u=1}^n C_{\e,\pset}^{(u)}},\label{eq:mu}
\end{align}
where $u$ goes over the set of URLs appearing during a given month and $C_{\e,\pset}^{(u)} = \sum_{\p\in\pset} C_{\e,\p}^{(u)}$ sums the counts over the corresponding buckets of $\pset$.

Similarly, we also characterize the (weighted) standard deviation for each engagement type $\e$ and combined category $\pset$ according to
\begin{align}
    \begin{array}{c}
    \text{Domain}\\\text{Ideology}
\end{array}~\sigma^I_{\e,\pset} &= \sqrt{\frac{\sum_u C_{\e,\pset}^{(u)} \left(i(u) - \mu^I_{\e,\pset}\right)^{2}}{\sum_u C_{\e, \pset}^{(u)}}},&
    \begin{array}{c}
    \text{Domain}\\\text{Quality}
\end{array}~\sigma^Q_{\e,\pset} &= \sqrt{\frac{\sum_u C_{\e,\pset}^{(u)} \left(q(u) - \mu^Q_{\e,\pset}\right)^{2}}{\sum_u C_{\e, \pset}^{(u)}}}.
\label{eq:w_avg}
\end{align}

Given that we are again working with ratios involving noisy denominators, it is important to carefully evaluate the signal-to-noise ratio (SNR) of the denominator and to determine the uncertainty intervals for each component of the ideology gap. We obtain $\mbox{SNR}$ values exceeding $1 \cdot 10^5$ and uncertainty intervals for the averages of both $\mathcal{C}$ and $\mathcal{L}$ on the order of $1 \cdot 10^{-3}$ (see the \textit{Supplementary Information 1B}).
We therefore can conclude that the metrics are statistically significant. The weighted standard deviations are substantial and vary over time, so we plot them separately when needed.

We define our metric for segregation in news consumption, which serves as a proxy for polarization during a given period, as the absolute difference between the average ideological leaning of URLs engaged by conservatives $\mathcal{C}$ and liberals $\mathcal{L}$ as
\begin{align}
    I_{\mbox{GAP}} & =\left| \mu_{\e,\mathcal{C}} - \mu_{\e,\mathcal{L}} \right|.
\label{eq:gap}
\end{align}

We also quantify the prevalence of content from low-quality sources, which can be indicative of a higher risk of misinformation exposure, by calculating the proportion of URLs with a content quality score lower than a threshold $T_\text{low}$~\cite{lin2023high}:
\begin{align}
P_{\text{low}} = \frac{\sum_{u=1}^{n} C^{(u)}_{\e} \cdot\left[q(u) \leq T_\text{low}\right]}{\sum_{u=1}^{n} C^{(u)}_{\mathtt{e}}}.
\label{eq:low}
\end{align}

\subsection{R5a. Findings Regarding Ideological Segregation}

\begin{figure*}[t!]
\centering 
\includegraphics[width=0.99\textwidth]{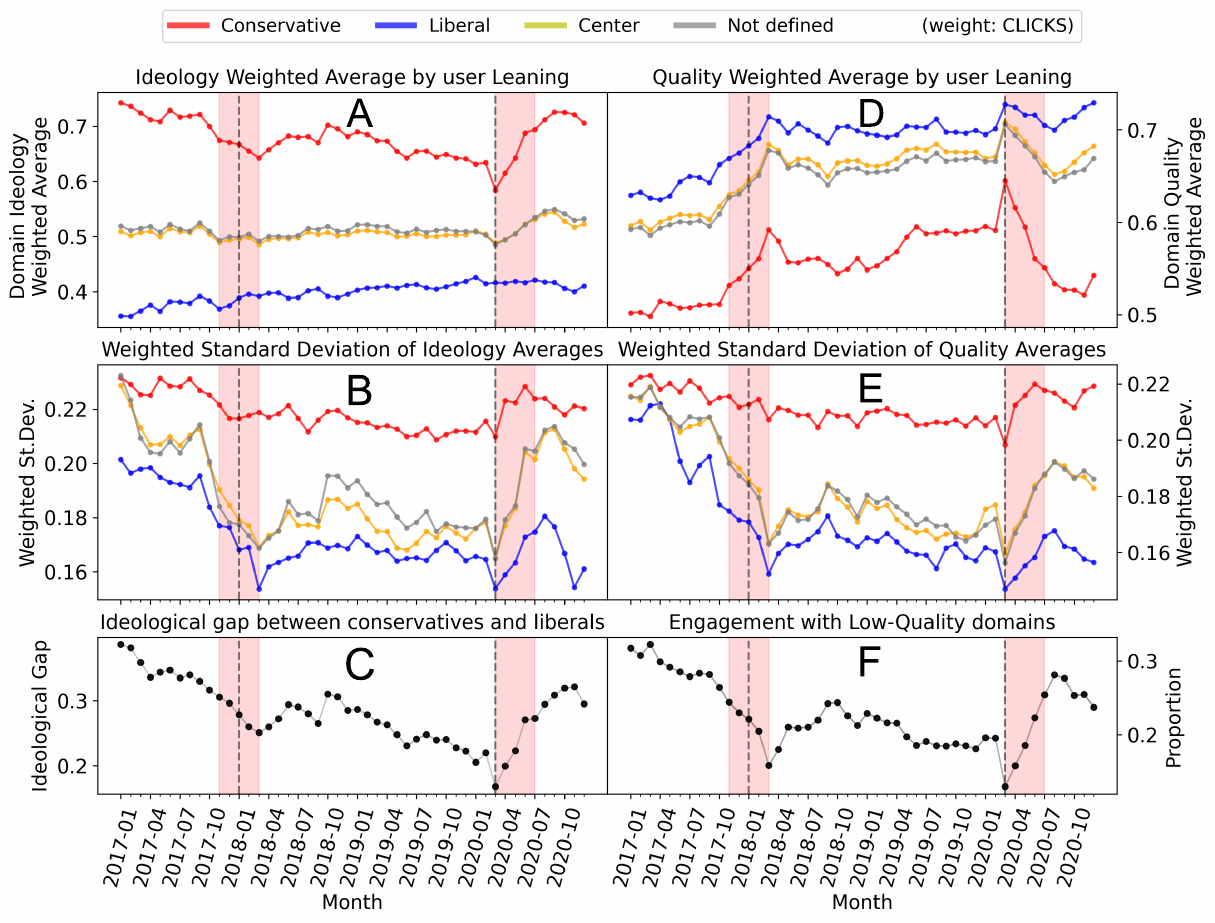} 
\caption{\textbf{Evolution of engagement (clicks) in terms of content ideology and content quality}. \textbf{(A)} and \textbf{(D)} show the evolution of the weighted averages of the domain ideology (normalized between 0 and 1) and domain quality, respectively, as in Eq~\eqref{eq:mu}. The averages are done for each user class (i.e. Conservative, Liberal, Centrist, or without a defined PPA) using clicks as weights. \textbf{(B)} and \textbf{(E)} show the weighted standard deviations related to the ideology and quality averages above, as in Eq~\eqref{eq:w_avg}. \textbf{(C)} represents the ideological gap between the conservative and liberal averages and can serve as a proxy for ideological segregation of news consumption, Eq~\eqref{eq:gap}. \textbf{(F)} shows the proportion of clicks directed towards low-quality domains, Eq~\eqref{eq:low} with $T_\text{low}=0.6$. Noise uncertainty intervals are too small to be visually detected in any of the subfigures. For all shown metrics that involve noisy denominators, we find $SNR > 10^5$. Moreover, bootstrapped uncertainty intervals for the weighted averages have very small standard deviations (in the order of 0.001), indicating high statistical precision. Dashed lines indicate relevant algorithmic updates. Shaded areas are calculated as in the previous Figure \ref{fig:engagement}. }
\label{fig:results}
\end{figure*}

To analyze the evolution of the ideology gap over time we compute the metrics introduced in equations~\eqref{eq:mu},\eqref{eq:w_avg} and~\eqref{eq:gap}.
Figure~\ref{fig:results} (left column) shows these results.
We observe in Figure~\ref{fig:results}A that, while the liberal and centrist weighted averages are stable and show a slight increase over time, the conservative counterpart shows larger variations with no stable trend, which include a noticeable increase after the second change point region. It is also interesting to note that centrist and undefined users show almost identical behavior. We present these results using clicks per user category as the weight of the averages but perform the same analysis also using different actions. There we find qualitatively similar results, even though conservative and liberal averages are positioned more to the extremes when considering shares and likes (\textit{Supplementary Fig. S6}). 

\begin{figure*}[t!]
\centering 
\includegraphics[width=0.99\textwidth]{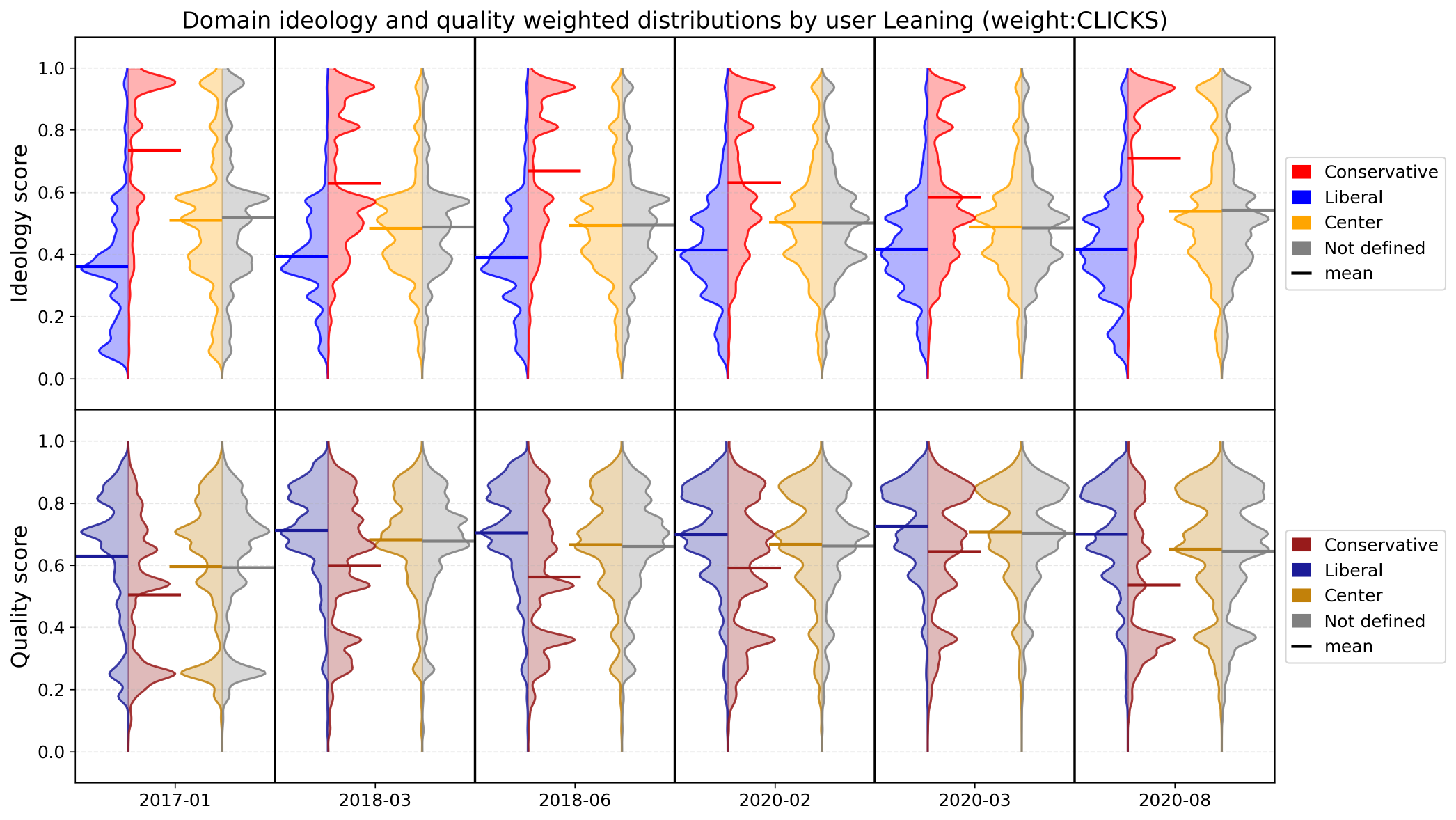} 
\caption{
\textbf{Distributions of weighted domain ideology and quality scores per user group at key time points.} Each violin plot represents the distribution of domain scores (ideology in the top row, quality in the bottom) obtained by applying the engagement-weighted distribution over URLs to the corresponding score function. Horizontal lines indicate the weighted average score for each user group: conservatives (red), liberals (blue), centrists (yellow), and undefined (gray). These averages correspond to the monthly values shown in panels A and D of Figure~\ref{fig:results}. The figure shows six selected months around the detected change points in engagement.
}
\label{fig:violin}
\end{figure*}

The weighted standard deviations (Figure~\ref{fig:results}B) are systematically higher for the conservative group, in accordance with the bimodal distribution of domains. In addition, we observe two noteworthy points: (i) Following the first shift in 2018, the weighted standard deviations of center- and liberal-leaning users decrease suddenly and change into a flat trend. (ii) In contrast, after the second shift in 2020, the weighted standard deviation for center-leaning users shows a sharp increase, potentially indicating that some individuals within this group began engaging with the extreme conservative group mentioned in the distribution of domains. 

Figure~\ref{fig:results}C shows the ideology gap between conservative and liberal groups (black line). We observe a gradual decline, except during the periods associated with the identified change points. In particular, in the second change point in 2020, the gap increases sharply. This rise can be attributed to a shift in the average consumption and engagement of conservative-leaning users toward more extreme-leaning domains, as illustrated in Figure~\ref{fig:results}A. In this period we observe a greater supply of more conservative content as well as an increased engagement with more extreme content from conservative users.

While our main analyses rely on weighted averages, we acknowledge that alternative summary measures, such as medians or full distributional comparisons, could yield different patterns. However, we verified that using the median instead of the mean results in similar overall trends, particularly regarding the persistent gap between conservative and liberal users and its variations around the algorithmic change points.
Moreover, to complement the trends shown in the time series, the top row of Figure~\ref{fig:violin} presents the full distributions (depicted as violin plots) of the weighted domain ideology scores for each user group at six key time points. Conservative users (in red) consistently exhibit a broader distribution than other groups. 

We observe that the initial decrease in the ideology gap is mainly explained by a shift in engagement of conservatives to less extreme right-leaning outlets, creating a noticeable concentration in the mid-right range (compare red distributions between the 1st and 2nd top plot). This decrease stops after the first change point. There is also a decrease in engagement in extreme left-leaning outlets (in blue).

The second change point is characterized by renewed engagement of conservatives with more ideologically extreme conservative domains (compare red distributions between the 5th and 6th top plots). This second shift is not exclusive to conservatives: it is also reflected among center (yellow) and undefined (gray) users, whose engagement distributions show a small but visible shift in mass toward more extreme right-wing outlets. In contrast, liberal users (blue) show a more stable distribution over time, with less structural change across the change points and less engagement with ideologically extreme sources.
\begin{figure}[!t]
\centering
\includegraphics[width=0.75\textwidth]{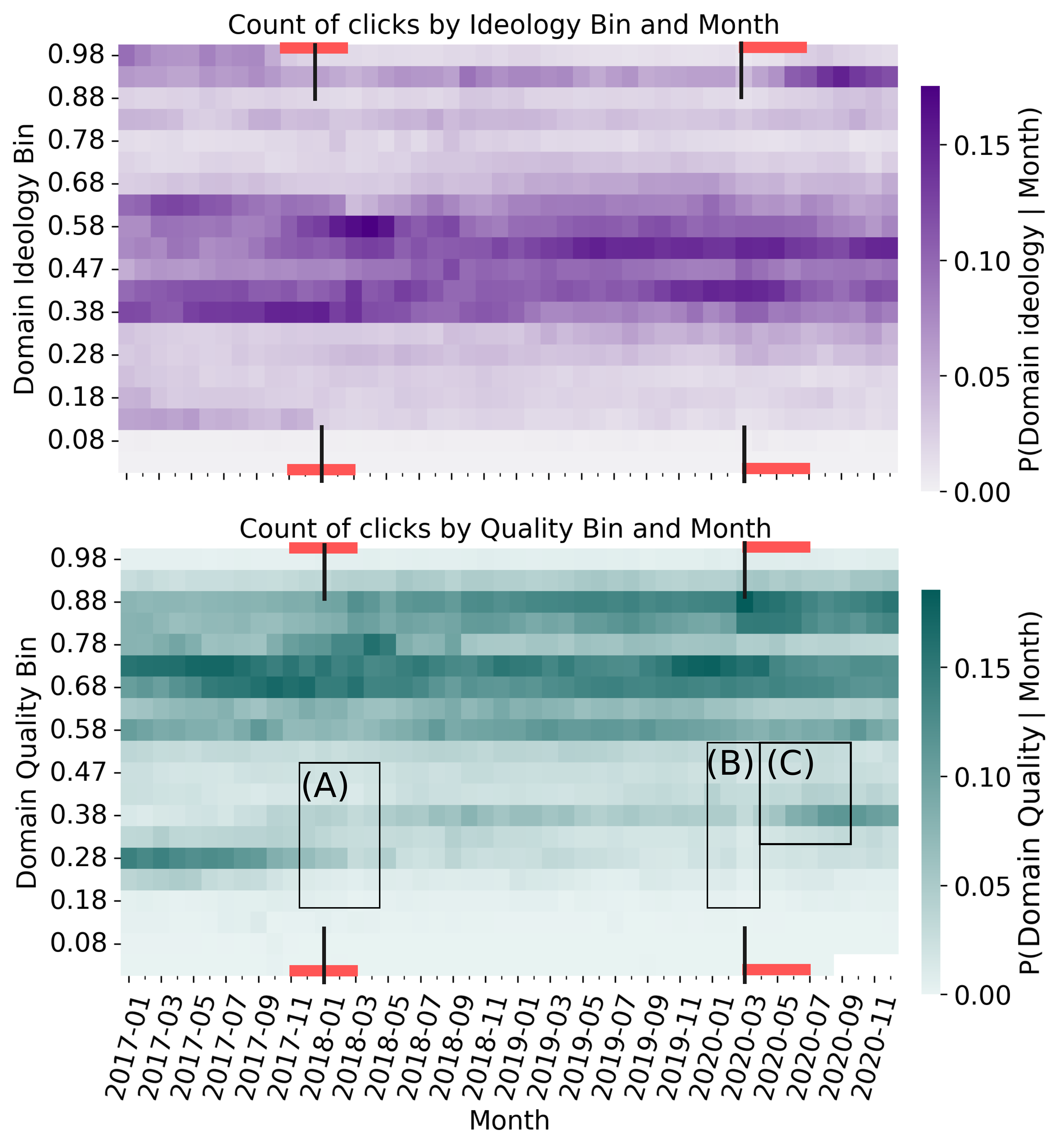} 
\caption{\textbf{Distribution of Clicks Over Time.} 
Relevant dates for algorithmic changes are marked with black line markers. Time intervals marked with a red marker indicate change point regions obtained through piece-wise linear regression. Ticks mark the center of the bin. \textbf{(top)} monthly distribution of clicks across \emph{domain ideology} bins (bin width $= 0.05$), normalized per month. \textbf{(bottom)}  monthly distribution of clicks across \emph{domain quality} bins (bin width $= 0.05$), normalized per month. A quality score threshold at $0.58$ separates high- and medium-quality domains from low-quality ones. For low-quality domains, we highlight three events (A), (B), and (C), which are associated with algorithmic change points.}
\label{fig:qtime}
\end{figure}

To close this subsection, the top panel of Figure~\ref{fig:qtime} illustrates the temporal evolution of ideological engagement by showing the monthly distribution of clicks across domain ideology bins. While engagement with non-extreme domains remains relatively stable, we identify two periods of increased engagement with domains close to the ideological center, visible as darker horizontal bands near the middle of the distribution. These periods align with the two local minima of the segregation measure shown in Figure~\ref{fig:results}C, supporting the trends described earlier in the subsection.

Finally, we remark that the increase in the segregation measure after the second change point, shown in Figure~\ref{fig:results}C, is clearly reflected in Figure~\ref{fig:qtime} as a sharp rise in clicks to far-right domains. This pattern is consistent with the rise in engagement with ideologically extreme conservative content shown in Figure~\ref{fig:violin}, especially among conservative and, to a lesser extent, undefined and center users.

\subsection{R5b. Findings Regarding Information Quality}

We now address \textbf{RQ5b}, focusing on the evolution of engagement trends for news-related URLs in relation to content quality.

Figure~\ref{fig:results}D shows the weighted averages of content quality consumed by different user classes, allowing us to investigate how these groups engage with varying content quality. There is a noticeable rise in the average consumed quality across all user classes prior to the second change point. 
Notably, during the periods associated with point changes, there is a sudden increase in average consumed quality across all user classes, which is followed quickly by a sharp decline. It is interesting that after the second change point region, liberal-leaning stays at higher values despite decreasing, while conservative-leaning goes back to lower quality.  Centrist and ideologically undefined users also show a bit of a decrease in their average quality as well as an increase in their weighted standard deviation (see Figure~\ref{fig:results}E), and thus may also be contributing to the higher engagement toward low-quality content.

In Figure~\ref{fig:results}F we show the change in the proportion of engagement directed at low-quality domains (quality bin lower than the bin centered at $0.58$ in Figure~\ref{fig:domains}), which fluctuates around $20\%$ and $30\%$. We observe two sudden decreases around change point regions, in the first case after the region (2018-03) and in the second case before the region (2020-03). We also see an immediate increase after the drops in both cases, even though it is steeper in the second region, achieving values similar to the start of the plot, with $~30\%$ of the engagement directed towards low-quality sources.
These two ``bounce-back'' effects are driven mainly by conservative-learning users. 
Note that the second bounce-back effect coincides with the Covid-19 pandemic and might be related to its respective information crisis.

Similar to the case of domain ideology, Figure~\ref{fig:violin} (bottom row) shows violin plots of domain quality distributions for each group for selected months around the main engagement change points. While all distributions are broad, the liberal (and centrist and undefined) groups exhibit a higher concentration of engagement in higher-quality domains compared to conservatives.

We observe two main changes in the low-quality region of domain quality scores which appear temporally related to the two minima of Figure~\ref{fig:results}F. The first, linked to the first change point, is characterized by the disappearance of a set of outlets with quality scores around $0.28$. This shift appears mainly in the center/undefined ideology user classes (compare 1st and 2nd bottom violin plots). The second change, associated with the second change point, is marked by a progressive increase in engagement with domains scoring around $0.38$, as shown in the last three panels of the bottom row.

In Figure~\ref{fig:qtime} (bottom) we show the distribution of clicks per month in each domain quality bin. The two distinct regions below and above the quality score bin centered at $0.58$ support our choice of a threshold around $0.6$ for trustworthy news, as recommended by NewsGuard~\cite{lin2023high}. The set of domains above this bin, which corresponds to medium and high-quality domains, dominates engagement (most of the clicks fall into that area of the figure), but there still are some low-quality domains that receive comparable engagement.

Regarding the trends in low-quality domains, we highlight three key events in Figure~\ref{fig:qtime} that help us better understand Figures~\ref{fig:results}F and ~\ref{fig:violin} (bottom row). The first event (A) coincides with the first algorithmic change and is characterized by a sudden drop in engagement with domains with quality scores around $0.28$. The second and third events occur during the period of the second algorithmic change: event (B) marks another sharp decline affecting all low-quality source domains, followed by event (C), a rapid rebound driven primarily by domains with quality scores around $0.38$.

\section{Discussion and Conclusions}

\subsection{Summary of findings}

The aggregate analysis shows U-shaped patterns of engagement that indicate more extreme users are more likely to engage actively with content, particularly through comments, likes, and expressions of anger (RQ1). 
This is in agreement with prior research showing that individuals with stronger ideological leanings and greater interest in news are more likely to participate in activities such as commenting and sharing~\cite{kalogeropoulos2017shares}. 

Additionally, we find that news diets of users, and engagement with such news, are biased toward their ideological leaning (RQ2). However, these biases are not symmetric. Consistent with observed trends of asymmetric polarization in congressional media engagement over time, driven primarily by a growing engagement extremity by Republicans in Congress \cite{heseltine2023asymmetric}, our findings reveal a pronounced bimodal pattern for conservative users,  characterized by a division between moderate and a more extreme sources. Liberals, on the other hand, have a shifted unimodal pattern. In terms of quality, we also find that conservative users predominantly consume content from lower-quality domains, with this tendency becoming more pronounced the more extreme the users are. This agrees with previous literature \cite{guess2021cracking,bailey2021interactions}.

Our findings align with prior research on the dissemination of vaccine-related news in Italy from 2016 to 2021, which also observed a U-shaped relationship between engagement and narrative bias, where more extreme narratives attracted higher engagement~\cite{galeazzi2024unveiling}. Our results provide additional evidence for this pattern, which can be understood as the outcome of multiple factors, including users’ tendency to engage predominantly with ideologically aligned news and the greater propensity of more extreme users to actively interact with news content.

The longitudinal analysis, on the other hand, reveals two distinct change points in engagement trends: an upward trend commencing in 2018 and a downward trend starting in 2020 (RQ3). We find the same results for the engagement to URLs that are not in our popular news domain list, which suggests this is a platform-wide pattern (RQ4). The initial upward trend is accompanied by a moderate increase in the ideological gap and a rise in the prevalence of lower-quality sources, which is followed by a gradual decline until early 2020. Conversely, the second change in the engagement trend is accompanied by a sizable increase in the ideological gap and prevalence of lower-quality sources followed by a moderate decrease later in 2020 (RQ5a and RQ5b). Since we observe two opposing changes in engagement (an increase in 2018 and a decrease in 2020) that lead to similar effects in terms of news partisan segregation and quality, we conclude that the relation between engagement and phenomena such as misinformation or ideological segregation is not trivial. 

Our analysis also identifies two distinct and abrupt decreases in the proportion of engagement with low-quality sources. Each of these declines is associated with the sudden disappearance of a different set of low-quality domains and are closely aligned with the detected change points (RQ5b).

Below, we connect these observations with their potential relation to known News Feed algorithm changes.

\subsection{Contextualization of our results with the Facebook News Feed algorithm}
\label{sec:context}

A major update in Facebook’s News Feed algorithm occurred at the start of 2018 \cite{meta_2018, wsj_2021, narayanan2023smalg}. This update had the purpose of overturning an observed downward trend in user engagement. It introduced the metric of Meaningful Social Interactions (MSIs) and rewarded posts that were likely to trigger certain types of engagement (especially comments and reshares, but also reactions such as \textit{haha’s}, \textit{wow’s} or \textit{angers}). A second update, somewhat less documented, appears to have occurred at the beginning of 2020 \cite{wsj_2021, narayanan2023smalg}. The purpose seems to have been not so much to boost engagement as to limit the spread of particularly toxic, divisive or, low-quality posts, through limiting long chains of reshares.  Among other things, it apparently took away the boost given to reshares, while keeping it high for comments \cite{wsj_2021, narayanan2023smalg}. 

A natural question that arises is whether the changes in engagement observed in the longitudinal analysis are related to the changes in the News Feed algorithm mentioned above and, eventually, what the underlying mechanism(s) may be. \citet{matias2023humans} and \citet{narayanan2023smalg} emphasize the importance of asking this kind of question.

Consider the 2018 update. It increased the weights attached to specific types of engagement (comments, shares, reactions etc.) with the stated objective of increasing overall engagement  (``meaningful social interactions'')  \cite{meta_2018,wsj_2021}. At the same time, however, as our longitudinal analysis suggests, the increase in engagement came with increased ideological segregation of news consumed by liberals and conservatives and a decrease in information quality. This would be in accordance with \citet{simchon2022troll}, whose results indicate that for political contexts polarized language is associated with higher engagement in social media. 

In terms of quality, we believe that a possible explanation for this phenomenon, based on the aggregate analysis, comes from the fact that more extreme users consume more extreme content and, particularly for conservative users, also of lower quality. This, combined with the U-shaped patterns of engagement, suggests that more extreme users are more likely to engage with comments, likes or \textit{angers} than moderate ones.
Hence an algorithmic update that boosts the weights on, say, comments (that are strongly U-shaped), makes posts that are likely to generate comments more visible to users. But because more extreme users are more likely to comment and such users tend to consume more extreme and (in part) also lower quality content, the algorithmic boost will tend to make more extreme and lower quality content more visible. The latter will in turn be more likely to be commented on, making it even more visible, and so on. Related to these questions, \citet{germano2022crowding} study such feedback loops occurring in response to algorithmic changes, and \citet{chavalarias2024can} also study associated network effects. 
Going through the feedback loop between algorithm and user engagement, applying the above logic to an update such as the MSI (2018), may well trigger increased engagement but also more ideological segregation in the distribution of that engagement (because of the more oppositely extreme content being consumed on both sides) and more low-quality information (because of the lower quality content being consumed, especially on the conservative side). 

By contrast, the 2020 update seems more difficult to reconcile with the observed patterns, since, unlike the 2018 update, it did not boost the weights on engagement but rather reduced them, especially for the reshares \cite{wsj_2021, narayanan2023smalg}. We observe reduced engagement but also an increased ideology gap and a stronger prevalence of low-quality sources. One point that may be worth emphasizing is that reshares have a visibly less U-shaped pattern than comments, whose weight remained high in the News Feed ranking algorithm. It is therefore not inconceivable that a reduction in the weight of reshares (not really U-shaped) while maintaining the weights on comments high (strongly U-shaped) may make comments relatively more important and may therefore boost those undesired effects in terms of consumption of more extreme and low-quality sources. Clearly, the precise role of the differences in weights for different engagement patterns and their relation to engagement as well as segregation of news consumption and diffusion of low-quality news is something that requires access to more detailed information and seems worth exploring further.

\section{Limitations}
We conclude by addressing the limitations of this study in Subsection~\ref{sec:study_issues}. Potentially related dataset issues and their implications are described in Subsection~\ref{sec:dataset_issues}.

\subsection{Study limitations}
\label{sec:study_issues}

This work focuses on interactions with some of the most engaged news-related URLs on Facebook from 2017 to 2021, specifically targeting content from the top 1\% of the most viewed news domains in the US during this period~\cite{buntain2023measuring}, which accounts for over $77\%$ of the total URL views on Facebook in the US from 2017 to 2020. Given the heavy-tailed distribution of engagement on social media platforms, this cutoff is not arbitrary but reflects a pragmatic balance between coverage and signal quality. Including low-engagement domains would introduce substantial noise due to the platform’s differential privacy constraints, as such domains typically hover near the dataset’s inclusion threshold and offer little usable signal. While it is possible to compare part of the analysis with other content from less popular or non-news sources, it is not feasible to meaningfully assess the ideological or quality bias of such content. Therefore, our approach provides critical insights into prominent trends in news consumption but may exclude niche content with limited reach. Consequently, the findings should not be generalized to less engaged, less circulated, or non-news content. 

Furthermore, because the data and engagement behaviors analyzed are specific to Facebook, our findings should not be generalized to other platforms or to the broader online information ecosystem without caution. Different social media platforms have distinct affordances, audiences, and content ranking mechanisms, which may shape ideological and quality-related engagement in different ways.

Another limitation of this study is that we use ideology and quality scores that are defined at the domain level rather than at the level of specific news URLs. This is well-documented, as using the average partisanship of source audiences provides a relative measure of partisanship~\cite{green2025curation, gonzalez2023asymmetric, buntain2023measuring}, rather than an absolute one. In our study, this distinction does not have significant implications for most domains. However, it could become relevant in domains where individual scores exhibit a multimodal distribution. This could result in slightly underestimated measures of the ideological gap, as well as slightly overestimated corresponding standard deviations.

Similarly, the domain-level quality scores used in this study are composite metrics that aggregate multiple dimensions (e.g. factual accuracy, transparency, and editorial standards) into a single reliability score~\cite{lin2023high}. While this provides extensive coverage, it limits interpretability, as a low score may reflect different underlying shortcomings across domains. As such, our analysis captures general trends regarding the reputation of domains but cannot disentangle specific quality-related mechanisms at the content level.

A related limitation of our study lies in its reliance on aggregate engagement data, which precludes us from analyzing individual-level media consumption behaviors. This raises the possibility that some of the trends we observe—particularly the growing ideological gap and engagement with low-quality content—may be driven by a relatively small group of highly active users. This concern has been raised by Guess~\cite{guess2021almost}, who found that while a small subset of users consistently engaged with highly partisan and ideologically skewed online media content, most individuals exhibit far more moderate and diverse media diets. Our results, while robust at the aggregate level, must therefore be interpreted with caution, as they may reflect the outsized influence of this hyper-engaged minority. This phenomenon of aggregation bias is an important consideration in the interpretive validity of our findings.

Moreover, our metrics not only do not make claims about individual-level ideological shifts, but also do not address affective polarization, or user attitudes toward opposing groups. Nor do we interpret our findings as direct evidence of increasing societal polarization. Rather, we study how user engagement behavior, which is shaped in part by algorithmic design choices, relates to the ideological distribution of news content consumed on the platform.

Finally, our study analyzes temporal patterns in engagement derived from time series data, but it is important to recognize that significant exogenous events, such as the COVID-19 pandemic (beginning in March 2020), the US 2018 Midterm elections (November 2018), and the US Presidential elections (November 2020), may have substantially influenced the observed engagement trends. While our discussion focuses on the association between engagement and the impact of modifications to Facebook's news feed algorithms, which are compatible with the reported phenomena, making causal statements remains challenging due to the concurrent influence of multiple external events on user exposure to news.

\subsection{Potential Dataset Issues and their Impact in this Study}
\label{sec:dataset_issues}

\noindent \textbf{\emph{The 100 shares cutoff}}---
The dataset only includes URLs that were shared publicly at least $100$ times a month (plus Laplacian noise) on Facebook~\cite{messing2020facebook}. While this censoring can generate bias~\cite{allen2021research}, 
it should not particularly affect our study, since we are already applying a filtering on news from popular domains~(see Subsection ~\ref{sec:domains}). However, this bias may skew towards viral posts within those domains, as well as a particular subset of users who engage with such content. Also, since we do not have access to individual-level data, we may be missing the effects of small population segments that may contribute to spreading low-quality information through reshares. 

\noindent \textbf{\emph{Impact of the Privacy Protection}}---
The Facebook URL dataset is protected with differential privacy~\cite{messing2020facebook, evans2023statistically} by perturbing URL counts with zero-centered Gaussian noise with variance depending on the interaction type, to prevent tracing actions to specific users or URLs.

Evans and King~\cite{evans2023statistically} show that ignoring such differential privacy protections can lead to unpredictable biased results. They provide corrections in feasible scenarios, especially for linear models. For example, the computation of uncertainty intervals in engagement counts is detailed in the \textit{Supplementary Information 1A}.
For more complex models that involve, e.g., weighted average metrics, there are no available corrections in closed form. For situations involving a noisy denominator, such as the ones in this work, Evans and King~\cite{evans2021statistically} and Buntain et al.~\cite{buntain2023measuring} show that for high enough signal-to-noise ratio estimates of the denominator, it is possible to ensure reliable metrics.

To assess the reliability of our results in the presence of noisy ratios, we take the following steps: (I) We compute the Signal-to-Noise Ratio (SNR) of the denominator~\cite{buntain2023measuring, evans2021statistically} to avoid problematic cases. A high value of SNR ensures that the noise is sufficiently low, preventing inflation of the metric or pathological outcomes such as having a zero in the denominator. Generally, for all metrics involving a noisy denominator in our analysis, we find SNR$~> 10^5$.
Although no universally accepted threshold exists, our SNRs are much larger than what has been considered sufficient in other studies. For example, an SNR threshold of $16$ was obtained by Buntain et al.~\cite{buntain2023measuring} for reliable ideology estimates. This suggests that our findings can also be considered reliable. (II) We determine the noise uncertainty intervals of the ratios using a worst-case scenario approach, as outlined in the \textit{Supplementary Information 1B}.

\noindent \textbf{\emph{Past issues with the Dataset}}---
The dataset has encountered occasional data integrity issues~\footnote{ See for example:  \hyperlink{https://www.nytimes.com/live/2020/2020-election-misinformation-distortions\#facebook-sent-flawed-data-to-misinformation-researchers}{https://www.nytimes.com/live/2020/2020-election-misinformation-distortions\#facebook-sent-flawed-data-to-misinformation-researchers}.}. A significant discrepancy was identified on September 10, 2021, where data from US-based users with undefined PPA, which account for more than half of the engagement in the platform, was missing. This issue was detected after researchers had already begun using the dataset and was reportedly retroactively corrected~\cite{allen2022addendum}. Such lapses highlight the challenges of relying on datasets managed by platforms with limited transparency in their data collection processes.

\section{Ethical considerations}

Access to the dataset is governed by an application process and formal agreement, facilitated through Facebook's servers to ensure user privacy. While this protects user data, it poses challenges for replication by other researchers and even ourselves, particularly if access is revoked or the dataset undergoes changes.

The dataset's limitations prevent us from disentangling the influence of algorithmic recommendations in user news feeds from the impact of their social network, considering both friend connections and pages. Friendship networks, and their degrees of homophily, play a role in the spreading dynamics of reshares in contrast to comments or likes. This complexity adds a layer of uncertainty to our analysis.

Finally, the opacity surrounding platform features, such as alterations in the weighting of engagement metrics in the recommendation pipeline and their role in the overall News Feed algorithm, poses a significant challenge. This limits our study to descriptive and exploratory analysis, leaving many questions regarding the platform's mechanisms as open questions or mild hypotheses.

We explicitly discourage using our research to draw potentially negative conclusions about subsets of users regarding their ideology.




\section*{Declarations}

\subsection{Supplementary Information}

This article has an accompanying supplementary file with additional information.

\subsection{Ethics approval and consent to participate}

This study was conducted under a Research Data Agreement (RDA) with Meta, which provided access to anonymized and aggregated data in compliance with applicable privacy laws and institutional research policies. No personally identifiable information was accessed or disclosed. The Institutional Committee for Ethical Review of Projects (CIREP) at Universitat Pompeu Fabra concluded that \textit{the dataset used in the research is to be considered anonymous data, according to the GDPR}.

\subsection{Availability of data and materials}

Facebook Data in this study were obtained from Meta, as part of Facebook Open Research \& Transparency (FORT), an initiative to facilitate the study of social media’s impact on society. 
Access to data and subsequent analysis was conducted on the Facebook Open Research and Transparency platform, approved through the Social Science One (SS1) process.
Researchers seeking permission to use the FORT Platform must 1)apply to become an approved partner and 2) sign the Research Data Agreement (RDA), a publicly available legal agreement. The RDA prohibits sharing Facebook Data with any third-party. Researchers may request access to Facebook Data at: \url{https://socialscience.one/rfps}. If a journal requests data to verify findings, the journal will also have to request access at: \url{https://socialscience.one/rfps}.

The code used to produce the results and plots within the data access platform is publicly available at \hyperlink{https://osf.io/}{https://osf.io/} with DOI 10.17605/OSF.IO/MQ9AK.

\subsection{Funding}

This publication is part of the action CNS2022-136178 financed by MCIN/AEI/10.13039/501100011033 and by the European Union Next Generation EU/PRTR. F. G.~acknowledges funding from Grant MICIU/AEI/UE-PID2023-153318NB-I00 and from the Spanish Agencia Estatal de Investigaci\'{o}n (AEI), through the Severo Ochoa Programme for Centres of Excellence in R\&D (Barcelona School of Economics CEX2019-000915-S). 

\subsection{Acknowledgements}

We would like to thank Social Science One and Meta for facilitating access to the Facebook URL Shares dataset for researchers, including our team. We also acknowledge the Meta Support Community for their support with the technical and logistical aspects of the data. Additionally, we appreciate the researchers who have previously worked with this dataset, whose insights and methodologies have set a valuable example, enabling us to further develop its applications and analyses.

\subsection{Authors' contribution}

E.F. was responsible for developing the code, generating the figures, conducting the analyses, and drafting the manuscript. F.G. contributed contextual insights and feedback regarding behavioral patterns, potential algorithmic changes and the discussion section. V.G. and A.K. offered continuous supervision and contributed original ideas. All authors participated in revising and finalizing the manuscript.

\bibliographystyle{unsrtnat}
\bibliography{references}  






\end{document}